\documentclass{article}
\usepackage{graphicx}
\usepackage{amsmath}
\usepackage{titlesec}
\usepackage{fancyhdr}
\usepackage{float}
\usepackage{mathrsfs}
\usepackage{indentfirst}
\usepackage{verbatim}
\titleformat{\section}{\bfseries\large}{\thesection}{1em}{}[]
\titleformat{\subsection}{\large}{\thesubsection}{0.5em}{}[]
\titleformat{\subsubsection}{}{\thesubsubsection}{0.5em}{}[]
\begin{document}
    \title{Transformer-Unet: Raw Image Processing with Unet}
    \author{Youyang Sha$^1$ Yonghong Zhang$^{1,2}$ Xuquan Ji$^1$ Lei Hu$^{1,2}$\\
    \\ \small{$^1$Beijing Zoezen Robot Co Ltd}
    \small{$^2$Beihang University}}
    \date{}
    \maketitle

    \fancyhead[L]{\small{Transformer-Unet: Raw Image Processing with Unet}}
    \fancyhead[R]{}
    
    \begin{abstract}
        Medical image segmentation have drawn massive attention as it is important in biomedical image
        analysis. Good segmentation results can assist doctors with their judgement and further improve
        patients' experience. Among many available pipelines in medical image analysis, Unet is one of the
        most popular neural networks as it keeps raw features by adding concatenation between encoder and
        decoder, which makes it still widely used in industrial field. In the mean time, as a popular model
        which dominates natural language process tasks, transformer is now introduced to computer vision tasks
        and have seen promising results in object detection, image classification and semantic segmentation
        tasks. Therefore, the combination of transformer and Unet is supposed to be more efficient than 
        both methods working individually. In this article, we propose Transformer-Unet by adding transformer
        modules in raw images instead of feature maps in Unet and test our network in CT82 datasets for 
        Pancreas segmentation accordingly. We form an end-to-end network and gain segmentation results better 
        than many previous Unet based algorithms in our experiment. We demonstrate our network and show our
        experimental results in this paper accordingly.
    \end{abstract}
    
    \section{Introduction}

    Convolutional neural networks(CNN), as a popular backbone for many computer vision tasks, have reached
    state-of-the-art efficiency in many areas. For segmentation tasks, many works based on CNN(Dai et al., 2015;
    Badrinarayanan et al, 2017) surpass previous methods based on statistic models such as Markov Random Fields(MRFs)
    and Conditional Random Fields(CRFs). However, both CNN and statistic models have their own edges since CNN can 
    capture regional features with shared parameters and sliding boxes and statistic models are solid for global 
    information as they establish connections between every pixel pair. So the combination of CNN and statistic models
    are extremely useful in segmentation as many previous works(Liu et al., 2018; Chen et al., 2018) have proven.

    Transformer was first proposed by Vaswani et al.(2017) and it soon became a popular model in natural language 
    process(NLP) tasks as it can pay attention to the whole sequence of array without losing any useful information
    like recurrent neural networks(RNN). Many works(Parmar et al., 2018; Carion et al., 2020) have introduced
    transformer to computer vision tasks by dividing pictures or feature maps into different parts and forming 
    sequences of array accordingly. These methods naturally perform better in capturing global features as transformer
    can establish relations between different arrays in a sequence.

    However, both models have their shortcomings as CNN is not sensetive to global features and transformer costs a
    lot of computational amount and cannot capture regional features efficiently. Dosovitskiy et al.(2021) further
    mention the importance of the size of datasets as transformer model is naturally hard to converge. Therefore,
    the combination of CNN and transformer is suited for many segmentation tasks(Zheng et al, 2021) as models 
    combining CNN and transformer can pay attention to regional and global features. 
    
    As an alternative choice for statistical models, transformer can improve the utilities of plain CNN. So we
    propose adding transformer layers in classical segmentation network Unet(Ronneberger et al., 2015) and forms 
    an end-to-end network w.r.t. semantic segmentation. Although previous work have combined transformer and Unet
    (Chen et al., 2021), we argue transformer can be a more straightforward part in Unet structure since previous
    work process feature maps with transformer instead of raw images. We argue that this may not be efficient since 
    pixel-wise relations have been formed by CNN and capturing relations between different patches can also be done
    by CNN with bigger kernel size in this case, which means this method does not highlight the advantage of transformer
    directly. Because raw images are important in medical image analysis, we propose
    processing inputs directly with transformer and decode its output step-by-step following Zheng et al.(2021). We 
    form our transformer layers like Dosovitskiy et al.(2021) and follow the design of Unet(Ronneberger et al., 2015)
    by concatenating feature maps from transformer and CNN in decoders. We name our network TUnet(Transformer and Unet)
    which surpass Unet, Attention Unet and TransUnet in CT82 datasets for Pancreas segmentation in our experiment. Our 
    model can run in a fast speed despite the massive size of transformer model based on multi-layer perceptron(MLP) and
    it can be trained easily in a computer with modern GPUs.
    
    \section{Related Work}
    
    \textbf{CNN} are extremely successful in computer vision(CV) tasks since the proposal of AlexNet(Krizhevsky et al., 2014)
    and they dominate machine learning models in domains such as object detection(Redmon et al., 2016) and human joint point
    detection(Cao et al., 2021). CNN models nowadays are very deep(Simonyan and Zisserman, 2015; He et al., 2016), they 
    usually contains millions of trainable parameters and multiple layers. Therefore, CNN models are typically more 
    efficient than statistical models such as Markov Random Fields(MRFs), resulting in better performance in many 
    areas. The majority of state-of-the-art algorithms in CV use CNN as their backbone and some of them(Zheng et al., 2015)
    also apply different network designs such as recurrent neural network(RNN).

    \textbf{Semantic Segmentation} task means signing labels to every pixel in an image. It majorly relies on CNN after Long
    et al.(2015) proposed fully convolutional networks. Two major types of work have been done in segmentation research.
    The first is proposing new methods for encoding and decoding(Chen et al., 2016; Noh et al., 2016; Wang et al., 2017). 
    These methods optimize feature capturing process in CNN and enable deep nerual networks understanding images better.
    The second is combining CNN with statistical models(Liu et al., 2016; Chen et al., 2018). These methods modify outputs
    of CNN for better predictions like transfer learning. Semantic segmentation have gained amazing results with these efforts
    and they are also helpful for object segmentation tasks(He et al., 2017).

    \textbf{Attention} modules were first introduced in NLP tasks as an alternative choice for RNN since self-attention can
    capture relationships between different parts of a sequence without losing information. Currently attention have been
    introduced in CV(Wang et al., 2017; Oktay et al., 2018) and it can extract suitable information together with CNN.
    As a model based on attention, Transformer was first proposed by Vaswani et al.(2017) as a major approach for machine 
    translation and it soon became a dominant model after the proposal of BERT(Devlin et al., 2018). Currently, it has been
    introduced as a new model for image recognition(Dosovitskiy et al., 2021), semantic segmentation(Zheng et al., 2021) 
    and many other CV tasks(Parmar et al., 2018; Carion et al., 2020). 

    Both CNN and transformer have their own edges so recent works such as TransUnet(Chen et al., 2021) apply them together by 
    building a CNN pipeline with the combination of transformer modules. These methods make CNN
    models more capable for global feature generation compared with plain CNN models. Some of transformer based neural
    network models have reached state-of-the-art efficiency in some datasets such as ViT(Dosovitskiy et al., 2021).
    These works suggest that transformer, together with its combination of CNN, has great potential in CV areas. Despite
    previous success of transformer and CNN, little models focus on how transformer can act in raw images since it is
    computationally costly. Many transformer-CNN models(Carion et al., 2020; Chen et al., 2021) only apply transformer in
    a feature maps as they are smaller in size and this may reduce computational amount. However, we believe this limit
    the performance of neural networks as relations between different patches is more clear in raw images so we apply 
    transformer in raw images directly. Our method gains positive results in our experiment.(see section 5.)

    \section{Method}

    Following Ronneberger et al.(2015), we first make a typical Unet as our CNN structure with bilinear interpolation
    as our up sample method and maximum pooling as our down sample method. For the convenience of implementation, we 
    design an almost symmetric network(see Section 5.2) which can be easily modified with attention modules
    (Oktay et al., 2018) and transformer modules(Chen et al., 2021). In TUnet however, the encoder and decoder do not 
    connect directly which is explained beneath in this section.
    \begin{figure}[t]
        \begin{center}   
            \includegraphics[scale=0.06]{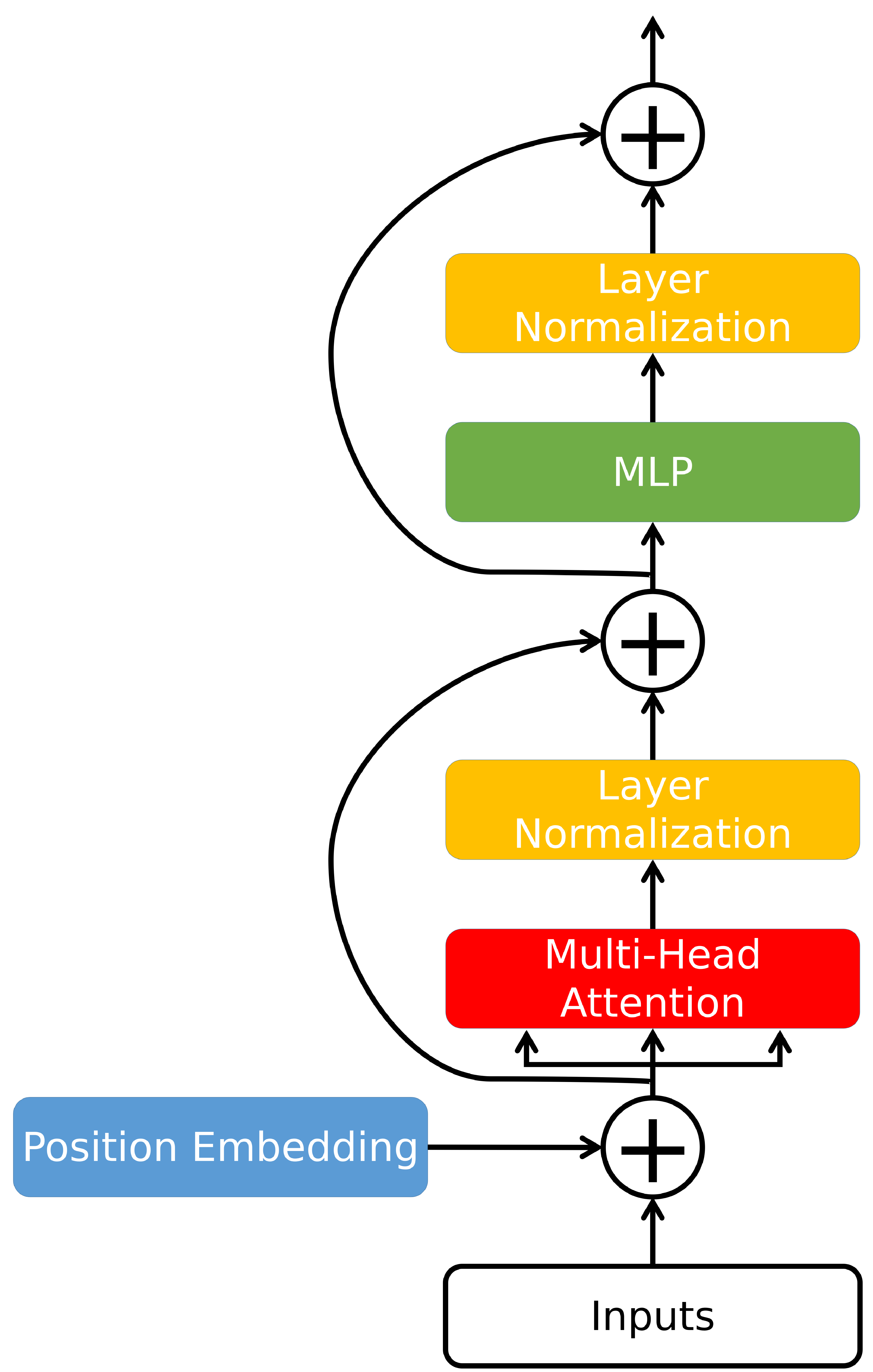}
            \includegraphics[scale=0.06]{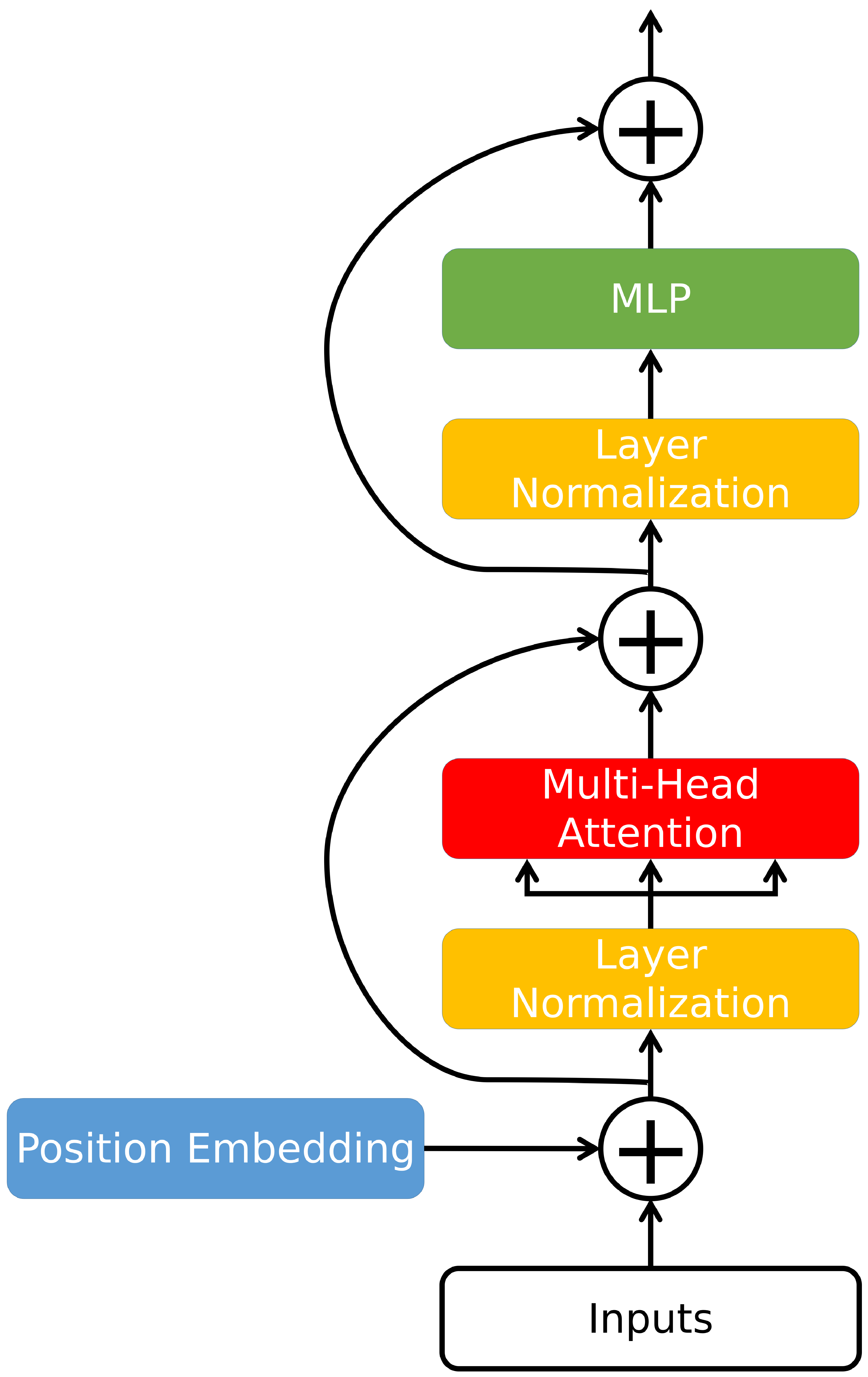}
            \caption{Transformer and vision transformer, the positions of layer normalization layers are different which makes
            ViT suits CV tasks better.}
        \end{center}
    \end{figure}

    As a model taking a sequence of arrays as input, transformer requires 1D data for segmentation tasks. Therefore, with
    a $C\times H \times W$ raw image, we flatten it into arrays of $C\times n^2$ dimensions where $n\times n$ is
    the size of image patches and $\frac{HW}{n^2}$ is the length of array sequences. We follow Dosovitskiy et al.(2021) by
    dividing a whole image into different squared pieces and $n$ is the length of square edges. To simplify the implementation
    process, in most cases we assume $H=W$ and $H,W$ can be divided evenly by $n$.

    \begin{figure}[t]
        \begin{center}
            \includegraphics[scale=0.08]{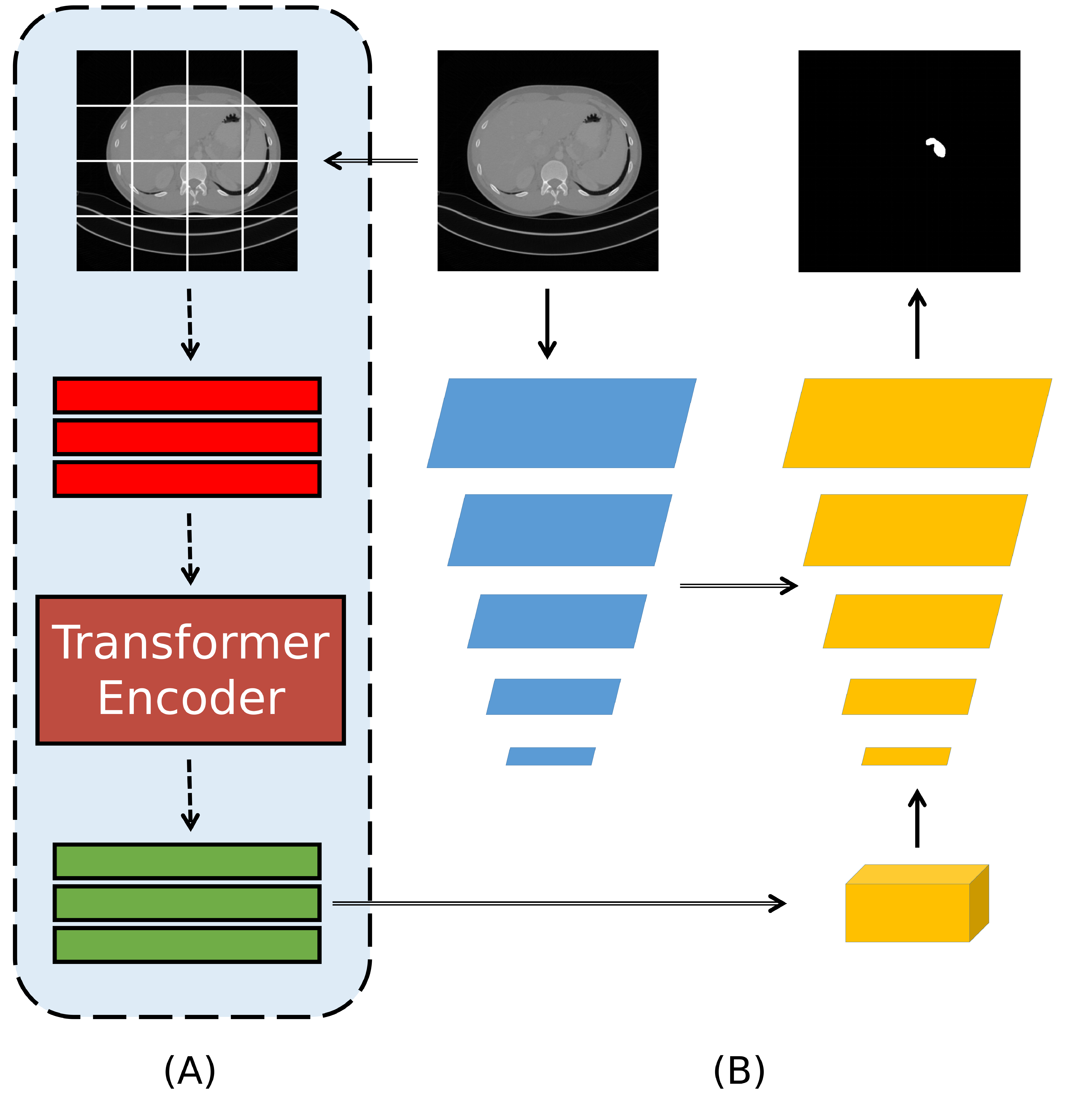}
            \caption{Our proposed network structure, part A is a transformer taking raw image as input and part B is a classical
            Unet where decoder takes the output of transformer as its input.}
        \end{center}
    \end{figure}

    We form our transformer model like Dosovitskiy et al.(2021), which is named vision transformer(ViT) and is slightly different
    from NLP transformer, as show in figure 1. ViT puts layer normalization before Multi-Head Attention and MLP as it ensures 
    the input values are not too big to process. In addition to that, ViT keep the major design of Vaswani et al.(2017) such as
    Multi-Head Self-Attention and MLP layers. Dosovitskiy et al.(2021) further add a learnable array for position embedding
    before feeding the whole sequence into transformer which is kept in our TUnet. We further modify ViT by replacing GELU
    with ELU as our activation function in MLP layers of transformer as we observe that ELU performs better in our experiment.
    ELU is seldomly used compared with RELU and GELU in transformer and it is defined as:
    \begin{equation}
        ELU=\begin{cases}
            x,if\  x\geq 0\\
            \alpha e^x-1,if\ x<0
        \end{cases}
    \end{equation}
    we argue that ELU is useful because negative values are as important as positive values in CT images. In our experiment, we 
    set hyper parameter $\alpha$ to 1.

    With methodology explained above, we form our transformer model with the following equations:
    \begin{equation}
        z_0 = [x_1E;x_2E;\cdots;x_nE]+E_{pos}
    \end{equation}
    \begin{equation}
        z'_l = MHA(LN(z_{l-1}))+z_{l-1}
    \end{equation}
    \begin{equation}
        z_l = MLP(LN(z'_l))+z'_l
    \end{equation}
    where MHA represents Multi-Head Attention layers, LN stands for layer normalization and $x_1,\cdots,x_n$ are image patches, 
    $l \in \{1,2,\cdots,m\}$ where $m$
    is the number of transformer layers. For raw image processing, we do the embedding process in ViT by applying a convolutional 
    layer with kernel size $1\times 1$ on the whole image, as $E$ in equation 2 indicates.

    Transformer is not as efficient as CNN in capturing regional features so we add an additional encoder following the design of 
    Unet(Ronneberger et al., 2015) in TUnet. This encoder does not connect with decoder directly. Instead, it outputs feature
    maps with different receptive fields and concatenate them with feature maps in decoder, as demonstrate in figure 2. Our
    decoder takes the output of transformer as its input, specifically, for transformer taking array sequences of size
    $\frac{HW}{n^2}\times Cn^2$, we reshape its output to size $C\frac{HW}{n^2}\times n\times n$ and feed it directly into decoder.
    By doing this, we ensure the input of decoder contains information of different image patches and is therefore better for
    our final prediction.

    \section{Implementation}

    Since we are processing raw images in TUnet, the sizes of original images and image patches are important because they determine
    the size of transformer model and its running speed. As we are choosing CT82 as our experimental datasets in which high resolution
    CT slices of size $512\times512$ include, we choose $16\times16$ as our image patches' size and therefore construct a sequence of
    length 1024. As a result, the input of decoder has a size of $1024\times16\times16$ in our experiment and we further reconstruct 
    it into size $1\times512\times512$ by bilinear interpolation. We add concatenation part in our decoder following Ronneberger
    et al.(2015) and build an encoder accordingly. To minimize our model while keeping its efficiency, the numbers of attention heads
    and total layers in our transformer module are 8 and 6.

    \section{Experiment}
    
    \subsection{Loss}

    To evaluate our model by comparing with other algorithms, we pick the most commonly used loss function in binary segmentation tasks,
    Binary Cross Entropy(BCE) loss, as our major optimization target. This loss function is simple and it cannot reflect the relations
    between pixels in our final predicted probability maps, therefore, it can better demonstrate how our model connect different parts
    of pictures. Generally, BCE loss is defined as:
    
    \begin{equation}
        L=\frac{1}{N}\sum_{i}-(y_i\cdot log(p_i)+(1-y_i)\cdot log(1-p_i))
    \end{equation}

    where $N$ is the number of pixels, $y_i$ is the label of pixel $i$, $p_i$ is the probability that pixel $i$'s label is true in our
    final prediction map. By definition is obviously that this function only calculates the loss of final prediction pixel-by-pixel 
    instead of regionally.

    \begin{figure}
        \begin{center}
            \includegraphics[scale=0.08]{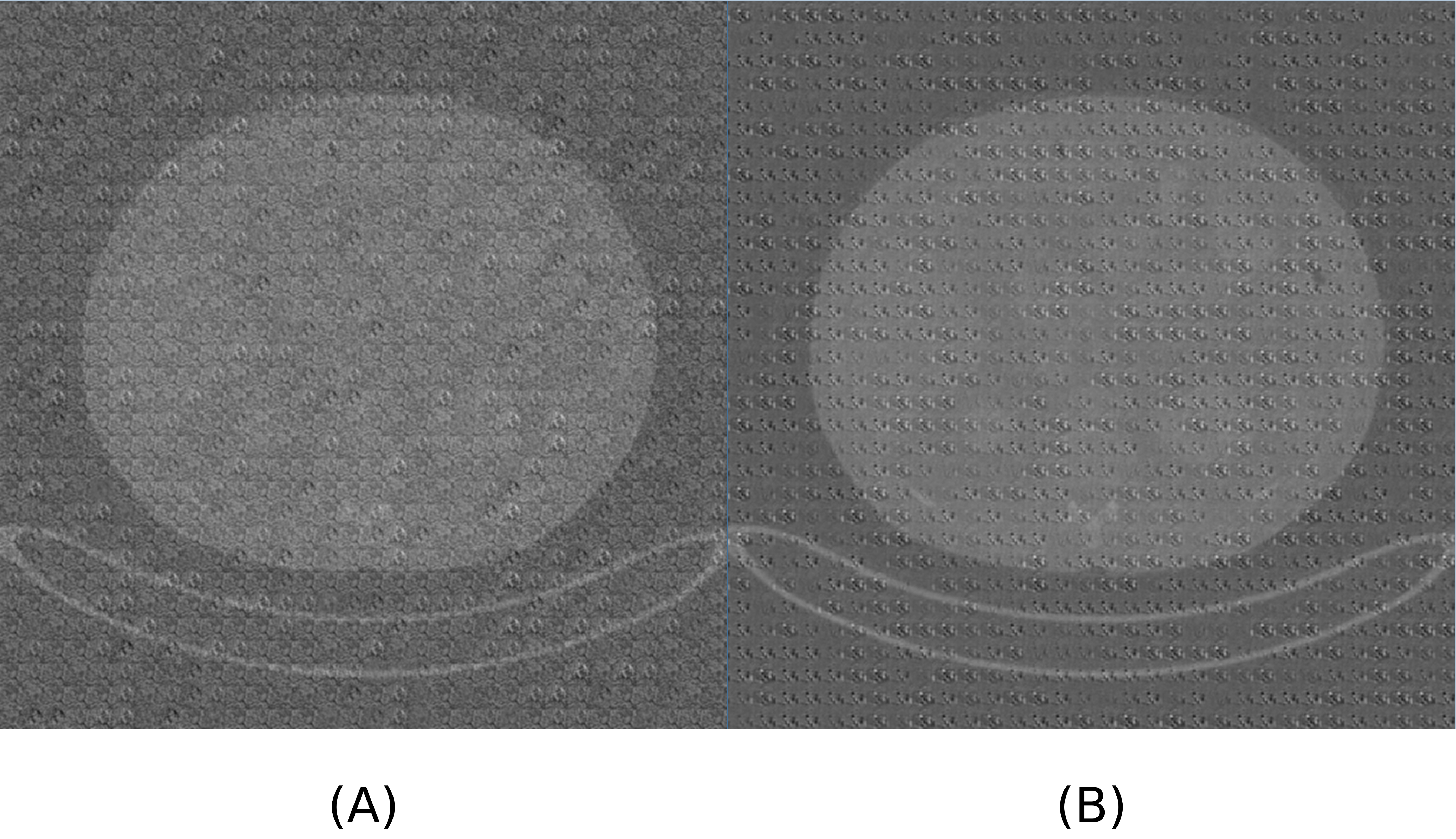}
            \caption{A is the output of first layer in transformer while B is the output of last layer. Transformer enables the
            network capturing global abstract features with a few layers. The background is the original image.}
        \end{center}
    \end{figure}

    \subsection{Setup}

    Previously, no work has processed raw biomedical images with transformer, so no pre-trained model is available for our CT82 datasets.
    Due to that, we replace the traditional pre-tranied and fine-tuning pattern with end-to-end training in the whole model. Another reason
    is that for TUnet, the major feature extractor is transformer module, so we design our Unet encoder shallower than decoder to decrease
    the size of our model and guarantee the performance of decoder. This is also why we call our model 'almost symmetric' in Section 3.
    The same backbone is also used in Attention Unet and TransUnet.

    We pick the developed self-adjusted optimizer AdamW(Loshchilov and Hutter, 2019) with an initial learning rate of $10^{-3}$. We train our 
    model for a total of 120 epochs and follow traditional stage training by decreasing our learning rate by a half after 60 and 100 epochs. 
    We also set the decay percentage rate to $10^{-6}$ to avoid overfitting and train our model with a single NVIDIA 3090 
    GPU(24GB graphic memory).

    The original CT82 datasets contains 3D CT files and for models like transformer, it is better to train with single CT slice in segmentation
    tasks. This is decided based on the following facts:

    \begin{itemize}
        \item The size of datasets is important for transformer(Dosovitskiy et al., 2021). By processing CT slices instead of the whole CT series, we
        are able to enlarge the size of datasets.
        \item Transformer cost a lot of graphic memory as it is based on MLP. So transformer is more suitable for 2D images because it does not increase
        the size of weight files massively.
    \end{itemize}

    We therefore process CT slices in our experiments and compare TUnet with existing models Unet, Attention Unet and TransUnet. To make our model
    better process data, we divided the whole image matrix with 1024, which is the approximate maximum absolute value of all CT slices in the datasets.

    \subsection{Results}
    \begin{table}[t]
        \begin{center}
            \begin{tabular}[t]{c|ccccc}
                \hline
                network & mIOU & Dice Score & Pixel Accuracy & Precision & Recall \\\hline
                Unet & 0.8113 & 0.7689 & 0.9981 & 0.8249 & 0.7200 \\ \hline
                Attn-Unet & 0.8172 & 0.7777 & 0.9982 & 0.8346 & 0.7280 \\ \hline
                TransUnet & 0.7882 & 0.7330 & 0.9979 & 0.8379 & 0.6515 \\ \hline
                TUnet & \textbf{0.8301} & \textbf{0.7966} & 0.9983 & 0.8278 & \textbf{0.7676} \\ \hline
            \end{tabular}
        \end{center}
        \caption{Utilities of networks based on different validation indexes}
    \end{table}
    \begin{table}
        \begin{center}
            \begin{tabular}[t]{c|cc}
                \hline
                network & Params & Inference Time \\\hline
                Unet & 506.6MB & 0.028s \\ \hline
                Attn-Unet & 508.0MB & 0.030s \\ \hline
                TransUnet & 1.2GB & 0.034s \\ \hline
                TUnet & 548.6MB & 0.041s \\ \hline
            \end{tabular}
        \end{center}
        \caption{Size and speed of different networks}
    \end{table}

    Our major evaluating method is multiple validation indexes including mIOU value and dice score of our final prediction. CT82 datasets is seperated
    to 60/22 for training and testing. The lowest resolution is $16\times 16$ in our model and this also applies to Unet, Attention Unet and TransUnet.

    For results demonstration, we set threshold to 0.8(i.e. a pixel with value larger than 0.8 in the final prediction map will be considered as a 
    Pancreas point) and consider not only the accuracy of Pancreas segmentation but also the recognition of background when calculating mIOU and pixel 
    accuracy values.

    Figure 3 shows a major advantage of transformer and this enables our model doing feature capturing job globally and locally with a few transformer
    layers. Table 1 shows the performance of Unet and its variances including our network. With deep Unet model as backbone, our model is able to surpass
    Unet and its related networks including the popular Attention Unet. Table 2 shows the size and inference time of different models and our model does
    not increase the size and inference speed by a lot.

    Figure 4 shows the visual results of different networks, our TUnet is able to make good segmentation results for long-distance pixel pairs because of
    transformer and is therefore better than other previous networks based on Unet.

    \begin{figure}[t]
        \begin{center}
            \includegraphics[scale=0.08]{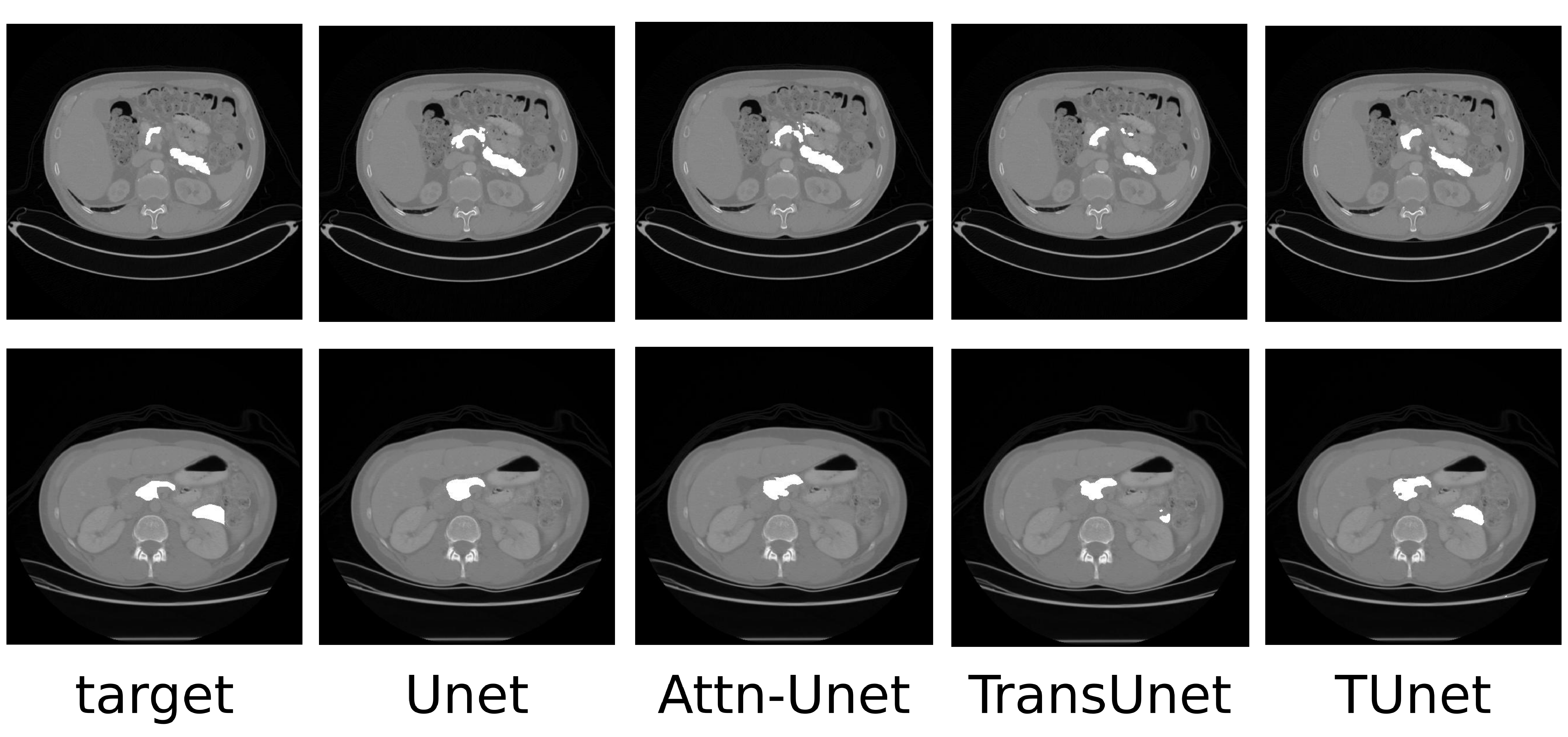}
            \caption{Visual results of different networks. Our positive results show that TUnet can make more precisive predictions when targets are
            far seperated.}
        \end{center}
    \end{figure}

    \subsection{Model Variance}

    In our experiment we choose $n=16$ as the size of image patches. However, many other options are available which suggest 16 may not be the
    ideal value for TUnet and we further perform experiments with $n=32,64$ respectively.

    Another important feature of TUnet is the deep and large Unet backbone. However, both Unet and Attention Unet are still useful in shallow models.
    Since deep models are not as convenient as shallow models because they naturally require better hardware such as GPUs, we further try our model
    with shallower Unet as backbone. In shallower model, we reduce $\frac{1}{3}$ CNN layers in Unet and decrease filter numbers to $\frac{1}{4}$ of
    deep model. The whole model is still end-to-end and we still train the raw model throughly. 

    \begin{table}[t]
        \begin{center}
            \begin{tabular}[t]{c|ccccc}
                \hline
                Resolution & mIOU & Dice Score & Pixel Accuracy & Precision & Recall \\\hline
                $16\times 16$ & \textbf{0.8301} & \textbf{0.7966} & 0.9983 & \textbf{0.8278} & \textbf{0.7676} \\ \hline
                $32\times 32$ & 0.8044 & 0.7584 & 0.9980 & 0.8140 & 0.7099 \\ \hline
                $64\times 64$ & 0.8008 & 0.7529 & 0.9980 & 0.8404 & 0.6818 \\ \hline
            \end{tabular}
        \end{center}
        \caption{Utilities of networks with various resolution based on different validation indexes}
    \end{table}

    \begin{table}[t]
        \begin{center}
            \begin{tabular}[t]{c|ccccc}
                \hline
                network & mIOU & Dice Score & Pixel Accuracy & Precision & Recall \\\hline
                Unet & 0.8011 & 0.7532 & 0.9980 & 0.8362 & 0.6853 \\ \hline
                Attn-Unet & 0.8112 & 0.7686 & 0.9981 & 0.8414 & \textbf{0.7075} \\ \hline
                TransUnet & 0.7894 & 0.7348 & 0.9979 & 0.8503 & 0.6470 \\ \hline
                TUnet & 0.8078 & 0.7634 & 0.9981 & \textbf{0.8639} & 0.6839 \\ \hline
            \end{tabular}
        \end{center}
        \caption{Utilities of networks with shallow Unet backbone based on different validation indexes}
    \end{table}

    Our results of variant models are listed in table 3 and table 4. From table 3 we can observe that for TUnet, $16\times 16$ is the optimal
    resolution for transformer and high resolution may weaken the efficiency for transformer since the length of array sequences is decreasing
    at the same time which is essential for self-attention layers in transformer. From table 4 we observe that TUnet shows no obvious advantage
    when using shallow Unet as backbone. Therefore the abstract features extracted by transformer may require deeper models for decoding. 

    \section{Conclusion}

    In this article we propose a transformer and Unet based nerual network for medical image analysis. The transformer process raw image directly
    instead of extracted feature maps. Our model is able to surpass other Unet based methods when applying deep backbones. However, this method
    does not improve Unet's efficiency a lot when using shallow models and we may apply more modification in the future for utilities improvement.

    \section*{Acknowledgement}

    This work is supported by Natural Science Foundation of Beijing, China (No.L202010).

    \newpage
    \thispagestyle{empty}
    
\end{document}